\documentclass[journal,draft,onecolumn]{IEEEtran}

\usepackage[final]{graphicx}
\usepackage[reqno]{amsmath}
\usepackage{amssymb}

\newfont{\bbb}{msbm10 scaled 500}

\newfont{\bb}{msbm10 scaled 1100}

\newcommand{\Prob}{\textrm{Pr}}

\newcommand{\hv}{{\bf h}}

\newcommand{\zerov}{{\bf 0}}

\newcommand{\Wc}{{\cal W}}

\newcommand{\Xc}{{\cal X}}
\newcommand{\Yc}{{\cal Y}}

\newtheorem{theorem}{Theorem}
\newtheorem{lemma}[theorem]{Lemma}
\newtheorem{definition}{Definition}

\author{Karim~Khalil, O.~Ozan~Koyluoglu, Hesham~El~Gamal, and Moustafa~Youssef}

\title{Opportunistic Secrecy with a \\Strict Delay Constraint
\thanks{This work is submitted to the IEEE Transactions on
Information Theory.}
\thanks{Karim Khalil and Moustafa Youssef are with the
Wireless Intelligent Networks Center (WINC),
Nile University, Cairo, Egypt.
Email: \{kareem.makarem,mayoussef\}@nileu.edu.eg.
O. Ozan Koyluoglu and Hesham El Gamal are with the
Department of Electrical and Computer Engineering,
The Ohio State University,
Columbus, OH 43210 USA.
Email: \{koyluogo,helgamal\}@ece.osu.edu.}
\thanks{This work is supported in part by an NPRP grant
from the QNRF, the Egyptian NTRA, and the National Science Foundation (NSF).
The material in this paper was presented in part at the
IEEE International Symposium on Information Theory,
Seoul, Korea in July 2009.}
}

\begin{document}
\maketitle


\begin{abstract}
We investigate the delay limited secrecy capacity of the flat fading
channel under two different assumptions on the available
transmitter channel state information (CSI). The first scenario assumes
perfect prior knowledge of both the main and eavesdropper channel gains.
Here, upper and lower bounds on the delay limited secrecy capacity
are derived, and shown to be tight in the high signal-to-noise ratio
(SNR) regime. In the second
scenario, only the main channel CSI is assumed to be available at the
transmitter where, remarkably, we establish the
achievability of a non-zero delay-limited secure rate, for a wide class of channel
distributions, with a high probability.
In the two cases, our achievability arguments are based on a novel
two-stage key-sharing approach that overcomes the {\em secrecy outage}
phenomenon observed in earlier works.
\end{abstract}


\section{Introduction}

Many wireless applications are limited by different 
forms of delay constraints. These applications
range from the most basic voice communication to the more
demanding multimedia streaming. However, due to
its broadcast nature, the wireless channel is vulnerable to
eavesdropping and other security threats. Therefore,
techniques that satisfy \emph{both} the delay limitation 
and the confidentiality requirement
are of definite interest. This motivates our analysis of the
fundamental (information theoretic) limits of secure
communication over fading channels subject to strict deadlines.

Recent works on information theoretic security have been largely
inspired by the wire-tap channel model of Wyner~\cite{Wyner:The75}.
In this seminal work, Wyner established the achievability of non-zero
secrecy capacity when the wiretapper channel is a
degraded version of the main one, by exploiting the noise to
create an advantage for the legitimate receiver. More recently,
the effect of fading on the secrecy capacity was studied
in~\cite{Gopala:On08} in the ergodic setting. The main insight
offered by this work is the achievability of a 
non-zero secrecy capacity, by {\em opportunistically}
exploiting the multi-path channel fluctuations,
even when the eavesdropper channel is better than the
legitimate one {\bf on the average}.

On the other side, delay limited transmission over fading channels
has been well studied in different network settings and using
various traffic models. For example, in~\cite{Hanly:Multiaccess:98},
the delay limited capacity notion was introduced and the optimal
power control policies were characterized in several interesting
scenarios. In~\cite{Berry:Communication02}, the strict delay
limitation of~\cite{Hanly:Multiaccess:98} was relaxed by allowing for
buffering the packets at the transmitter. In this setup, the asymptotic
behavior of the power-delay trade-off curve was characterized yielding
valuable insights on the structure of the optimal resource allocation
strategies~\cite{Berry:Communication02}. More recently, the scheduling
problem of data transmission over a finite delay horizon assuming
perfect CSI was considered~in~\cite{Lee:Energy-efficient}. Our work can
be viewed as a generalization of~\cite{Hanly:Multiaccess:98} whereby a
secrecy constraint is imposed on the problem.

The delay limited transmission of secure data over fading channels
was considered previously in~\cite{Bloch:Wireless08}. In that work,
the authors attempted to send the secure information using binning
techniques inspired by the wiretap channel results. The drawback
of this approach is that it fails to secure the information in
the particular instants where the eavesdropper channel gain is
larger than that of the main channel. This results in the so-called
{\bf secrecy outage} phenomenon (as defined in \cite{Bloch:Wireless08}).
Unfortunately, in the delay limited setting, the secrecy outage
can not be made to vanish by increasing the transmission power, since
it does not offer a relative advantage to the legitimate receiver,
leading to the conclusion that the delay limited secure rate
achieved by this approach is equal to zero for most channel
distributions of interest~\cite{Bloch:Wireless08}.
This obstacle is overcome by our two-stage approach. Here, the
delay sensitive data is secured via Vernam's one time pad
approach~\cite{Vernam:OneTimePad26} 
(see also~\cite{Shannon:Communication49}) using
a private key, which was shared {\bf secretly} by
the two legitimate nodes during previous transmissions.
Since the key packets are {\bf not delay sensitive}, the two nodes
can share the key by distributing its bits over many fading
realizations to capitalize on the ergodic behavior
of the channel. Our result is enabled by observing that,
through the appropriate rate allocation, the key bits can be
{\bf superimposed} on the delay sensitive data packets so that
they can be used for securing future packets.
This mechanism is referred as
{\em key renewal} process in the sequel. This process requires an
initialization phase to share the key needed for securing the
first data packets. However, the loss in throughput entailed by
the initialization overhead vanishes in the asymptotic limit of
a large number of data packets. Our analytical results establish
the asymptotic optimality, with high SNR, of this novel approach
in the scenario where both the main and eavesdropper channel
gains are known {\em a-priori} at the transmitter.
When only the main channel
CSI is available, this approach is shown to achieve
{\bf a non-zero constant} secure rate for a wide class of
{\em quasi-static and invertible channels}~\cite{Hanly:Multiaccess:98}
with high probability~\footnote{We use a modified version of the
$\epsilon$-achievable rate defined in \cite{Caire:Optimum99}
to argue that our results are achievable with a vanishing
probability of secrecy outage.}.

The rest of the paper is organized as follows. Section~II
introduces the system model and notations used throughout the paper.
Section~III focuses on the full CSI scenario whereas the
case with only the main channel CSI is analyzed in Section~IV;
along with some representative numerical results. Finally,
some concluding remarks are offered in Section~V.


\section{System Model}

\begin{figure}[thb]
    \centering
    \includegraphics[width=0.6\columnwidth]{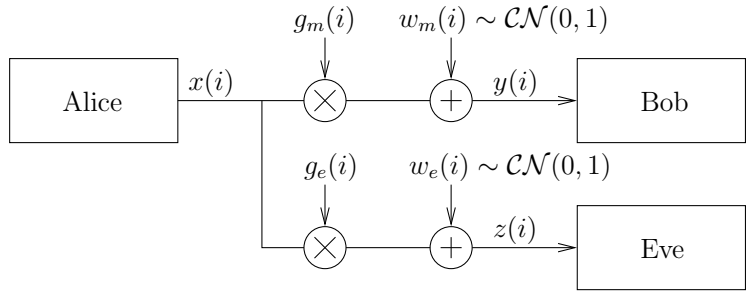}
    \caption{System Model.}
    \label{fig:model}
\end{figure}

The system model is shown in Figure~\ref{fig:model}. A source node
(Alice) wishes to communicate with a destination node (Bob) over a fading
channel in the presence of an eavesdropper (Eve). We adopt a block
fading model, in which the channel is assumed to be constant during a
coherence interval and changes randomly from an interval to the next
according to a bounded continuous distribution. Also, the coherence
intervals are assumed to be large enough to allow for the use of
random coding arguments. During any symbol interval 
$i$, the signals received at the destination and 
the eavesdropper, respectively, are given by
\begin{eqnarray}
y(i)&=&g_m(i)\, x(i)+w_m(i),\\
z(i)&=&g_e(i) x(i)+w_e(i),
\end{eqnarray}
where $x(i)$ is the transmitted symbol, $g_m(i)$ and $g_e(i)$ are
the main channel and the eavesdropper channel gains
respectively, $w_m(i)$ and $w_e(i)$ are the i.i.d. additive white
complex gaussian noise with unit variance at the legitimate receiver
and the eavesdropper, respectively. We denote the power gains
of the fading channels for the main and eavesdropper channels
by $h_m(i) = |g_m(i)|^2$ and $h_e(i) = |g_e(i)|^2$, respectively.
We impose the long term average power constraint $\bar{P}$, i.e.,
\begin{eqnarray}
{\mathbb E} [P(\hv)]\leq \bar{P},
\end{eqnarray}
where $P(\hv)$ is the power allocated for the channel state
$\hv=(h_m,h_e)$ and the expectation is over the channel gains.

The source wishes to send a message $W \in \Wc = \{1,2,\cdots ,M\}$
to the destination while satisfying the delay and secrecy
constraints. In the following, our delay
constraint is imposed by breaking our message
into packets of equal sizes, where each one is encoded
independently, transmitted in only one coherence block,
and decoded by the main receiver at the end of this block.
The total transmission time of $n$ channel symbol durations
is divided into coherence intervals of length
$n'$ channel uses; for which both $h_m$ and $h_e$ remain fixed. 
We assume there are total of $SB$ number of such
fading blocks. These blocks are grouped into
$S$ super-blocks, each consisting of $B$ fading blocks.
We will further represent a fading block with tuple
$(s,b)$ such that $s \in \{1,2,\cdots,S\}$ denotes the
super-block index and $b \in \{1,2,\cdots,B\}$ denotes
the fading block index within a particular super-block.
We consider the problem of constructing $(M',n')$
codes ($M=S\: B\: M'$) to transmit
the message of the block $(s,b)$, which is represented by
$W(s,b)\in \Wc' = \{1,2, \dots, M'\}$, to the receiver.
Here, an $(M',n')$ code consists of the following
elements: 1) a stochastic encoder $f_{n'}(.)$ at the
source that maps the message $w(s,b)$ to a codeword
$X^{n'}(s,b) \in \Xc ^{n'}$, and 2) a decoding
function $\phi$: $\Yc^{n^*} \rightarrow \Wc' $
at the legitimate receiver, where $n^*= (s-1)Bn'+bn'$
denotes the total number of the received signal dimension at
the receiver at the end of the block $(s,b)$. The average error
probability of an $(M',n')$ code is defined as
\begin{eqnarray}
P_e^{n'}(s,b)=\frac{1}{M'}
\sum \limits_{w \in \Wc'}
\Prob \left( \left\{\phi (y^{n^*})\neq w
| w \mbox{  is sent in block } (s,b) 
\right\}\right), \nonumber
\end{eqnarray}
where $y^{n^*}$ represents the total received signals at the
legitimate receiver at the end of the block $(s,b)$.
We define the equivocation rate $R_e$ at the eavesdropper
as the entropy rate of the transmitted message over block $(s,b)$
conditioned on the available CSI and all the channel outputs
at the eavesdropper, i.e.,
\begin{eqnarray}
R_e(s,b) \overset{\Delta}{=}
\frac{1}{n'} H(W(s,b)|Z^{n},h_m^n,h_e^n),
\end{eqnarray}
where $h_m^n=\{ h_m(1), \cdots, h_m(n)\}$
and $h_e^n = \{ h_e(1), \cdots, h_e(n)\}$ denote
the channel power gains of the legitimate receiver and the
eavesdropper in $n$ symbol intervals, respectively. We consider
only the perfect secrecy (in the sense of~\cite{Wyner:The75})
which requires the equivocation rate $R_e$ to be arbitrarily
close to the message rate. Hence, we define the achievability
of the delay limited secrecy rate and capacity, respectively,
as follows.
\begin{definition}\label{def:achievablerate}
The rate $R_{s,d}$ is said to be
\emph{an achievable delay limited secrecy rate},
if for any $\epsilon'>0$, there exist a
$(2^{n'R_{s,d}},n')$ code such that
\begin{eqnarray}
P_e^{n'}(s,b) &\leq& \epsilon', \nonumber \\
R_e(s,b) &\geq& R_{s,d}-\epsilon' \label{eq:secrecyCondition}
\end{eqnarray}
for every fading block $(s,b)$, $s\neq 1$, and for sufficiently large $n,B$.
The delay limited secrecy capacity, $C_{s,d}$, is defined as
the supremum of the achievable delay limited perfect secrecy
rates.
\end{definition}

Here, if the secrecy constraint \eqref{eq:secrecyCondition}
is not satisfied for a given block, then the corresponding
block is said to be in \emph{secrecy outage}, the probability
of which is defined as follows.
\begin{definition}
For a given $\epsilon'>0$, the probability of secrecy
outage for the block $(s,b)$ evaluated at rate $R_{s,d}$
is given by
\begin{equation}
P_{out}(s,b,R_{s,d},\epsilon') \triangleq  \Prob
\left( \left\{ R_e(s,b) \leq R_{s,d}-\epsilon' \right\} \right).
\end{equation}
\end{definition}

Now, we define a modified version of the
$\epsilon$-achievable rate notion given by~\cite{Caire:Optimum99}
(see also~\cite{Ozarow:Information94}) for the secrecy
outage phenomenon.
\begin{definition}\label{def:epsachievablerate}
The rate $R_{s,d}(\epsilon)$ is said to be
\emph{an $\epsilon$-achievable delay limited secrecy rate},
if for any $\epsilon'>0$,
there exists a $(2^{n'R_{s,d}(\epsilon)},n')$ code such that
\begin{eqnarray}
P_e^{n'}(s,b) &\leq& \epsilon' \nonumber\\
P_{out}(s,b,R_{s,d}(\epsilon),\epsilon') &\leq& \epsilon
\end{eqnarray}
for every fading block $(s,b)$, $s\neq 1$, and for sufficiently large $n,B$.
The $\epsilon$-delay limited secrecy capacity, $C_{s,d}(\epsilon)$,
is defined as the supremum of the achievable delay limited
secrecy rates with secrecy outage probability less
than $\epsilon$.
\end{definition}

We note that in our achievability
results, an initialization phase occurs during
the first super-fading block ($s=1$), and its
duration is negligible as $S\to\infty$. This explains why the 
requirements of Definitions \ref{def:achievablerate}
and \ref{def:epsachievablerate} are satisfied for
every fading block $(s,b)$ with $s\neq1$.

Finally, we give some notational remarks.
We denote the delay limited secrecy rate and capacity as
$R_{s,d}^{F}$, $C_{s,d}^{F}$, respectively, for
the full CSI scenario, where both $g_m$ and $g_e$ are
known {\em a-priori} at the transmitter.
For the main CSI scenario, where only $g_m$ is known
{\em a-priori} at the transmitter, the delay limited
secrecy rate, secrecy capacity, $\epsilon$-achievable secrecy
rate, and $\epsilon$-secrecy capacity, are denoted
respectively by $R_{s,d}^{M}$, $C_{s,d}^{M}$,
$R_{s,d}^{M}(\epsilon)$ and $C_{s,d}^{M}(\epsilon)$.
We let $[x]^+ = \max\{x,0\}$. $\log(\cdot)$ denotes
the base-2 logarithm. Throughout the sequel, the expectations
are taken with respect to the random channel gains.


\section{Full Transmitter CSI}
First, we give a simple upper bound on the delay limited secrecy capacity.
This bound will be used to establish the optimality of the
proposed two-stage approach in the high SNR regime.

\begin{theorem}\label{thm:upperFullCSI}

The delay limited secrecy capacity when both $g_m$ and $g_e$
are available at the transmitter, $C_{s,d}^{F}$, is upper bounded by
\begin{eqnarray}\label{eq:upperfull}
C_{s,d}^{F} \leq
\max \limits_{{}^{\quad \: P(\hv)}_{\textrm{s.t. }
{\mathbb E} [P(\hv)]\leq \bar{P} }}
\min
\left\{  R_s^{F},  R_d^{F}  \right\},
\end{eqnarray}
where $R_s^{F}$ and  $R_d^{F}$ are given as follows.
\begin{eqnarray}
R_s^{F} &=& \mathbb{E} \left[\log(1+P(\hv)h_m)-
\log(1+P(\hv)h_e)\right]^+\nonumber\\
R_d^{F} &=& \min\limits_{\hv} \log(1 + P(\hv)h_m )\nonumber
\end{eqnarray}
\end{theorem}

\begin{IEEEproof}
Consider an arbitrary power allocation scheme $P(\hv)$. 
Since imposing delay constraint
can only degrade the performance, we upper bound 
the achievable delay limited secrecy rate with 
the ergodic secrecy rate as
\begin{eqnarray}\label{eq:thm1eq1}
R_{s,d} ^{F} \leq R_s^{F}.
\end{eqnarray}
We also have
\begin{eqnarray}\label{eq:thm1eq2}
R_{s,d} ^{F} \leq R_d^{F},
\end{eqnarray}
since imposing the secrecy constraint can not
increase the achievable rate. Then,
combining \eqref{eq:thm1eq1} and \eqref{eq:thm1eq2},
and maximizing over $P(\hv)$, we obtain
\begin{eqnarray}\label{eq:thm1eq3}
R_{s,d} ^{F} \leq \max\limits_{P(\hv)}
\min \{ R_d^{F}, R_s^{F}\},
\end{eqnarray}
which proves our claim.
\end{IEEEproof}

The following result establishes a lower bound on the delay
limited secrecy capacity using our novel two-stage approach. The
key idea is to share a private key between Alice and Bob, without
being constrained by the delay limitation. This key is, then, used
to secure the delay sensitive data while {\bf overcoming the secrecy outage
phenomenon}. In the steady state, the key renewal process takes place
by superimposing the key on the delay sensitive traffic. More
precisely, as outlined in the proof, the delay sensitive traffic
(secured by the previously shared key) serves as a {\em randomization} signal
in the binning scheme used to secure the current key. Finally,
since $h_e$ is known {\em a-priori} at the transmitter, one
can further increase the delay limited secrecy rate by dedicating
a portion of the secure rate to the delay sensitive traffic
(as controlled by the function $q(\hv)$ in the following theorem).

\begin{theorem}\label{thm:lowerFullCSI}

The delay limited secrecy capacity in the full CSI
scenario, $C_{s,d}^{F}$, is lower bounded by the
following achievable rate.
\begin{eqnarray}\label{rates-1}
C_{s,d} ^{F} \geq R_{s,d}^{F}
= \max \limits_{{}^{P(\hv), \: q(\hv)}_{\textrm{s.t. }
{\mathbb E} [P(\hv)]\leq \bar{P} }}
\bigg[
\min \limits_{\hv}
\left\{
R_1(\hv) + R_2(\hv)
\right\}
\bigg],
\end{eqnarray}
where
\begin{eqnarray}\label{rates-2}
R_s(\hv) &=& \left[
\log(1+P(\hv)h_m)- \log(1+P(\hv)h_e)
\right]^+,\nonumber \\
R_k(\hv) &=&  [\log(1+P(\hv)h_m)-
\log(1+P(\hv)q(\hv))]^+ ,\\
R_2(\hv)
&=& R_s(\hv) - R_k(\hv), \nonumber
\end{eqnarray}
$q(\hv) \geq h_e$,  $\forall h_e$, and
$R_1(\hv)$ is chosen to satisfy the following
\begin{eqnarray}\label{rates-3}
\mathbb{E}[R_1(\hv)] &\leq& \mathbb{E}[R_k(\hv)]\nonumber\\
R_1(\hv) &\leq& \min \left\{\log(1+P(\hv)h_m),\log(1+P(\hv)h_e)\right\}
\end{eqnarray}

\end{theorem}

\begin{IEEEproof} \label{thm:lowerFullCSIproof}
Consider a fixed pair $(h_m,h_e)$, a power control policy $P(\hv)$ satisfying
$E[P(\hv)] \leq \bar{P}$, and an arbitrary function $q(\cdot)$
such that $q(\hv)\geq h_e$. The achievable rate is
obtained by finding the minimum rate over the pair $(h_m,h_e)$, to satisfy
our strict delay constraint, and then maximizing over all power
control policies and functions $q(\hv)$. We start the proof
by defining the different rates in 
(\ref{rates-1}),~(\ref{rates-2}),~(\ref{rates-3}): 
$R_s(\hv)$ is the instantaneous secrecy rate
supported by the channel, $R_k(\hv)$ is the rate used to share the private key,
$R_2(\hv)$ is the delay limited secrecy rate of the data that is
transmitted without the key, and $R_1(\hv)$ is the rate of 
the data sent via the one time pad
scheme. Moreover, we define the additional randomization rate by
\begin{eqnarray}
R_x(\hv) = \min \left\{\log(1+P(\hv)h_m),
\log(1+P(\hv)h_e)\right\} - R_1(\hv) \label{eq:thm2prfeq2}
\end{eqnarray}

{\bf Our Two-stage Scheme:} We divide the message
$W \in \Wc = \{1,2,\cdots,2^{nR_{s,d}^{F}}\}$ into
$(S-1)B$ data packets $D(s,b)$, each encoded independently
and sent with rate $R_{s,d}^{F}$ during the block
of the channel where $s \in \{2,\cdots,S\}$ and
$b \in \{1,2,\cdots,B\}$. We further divide each
data packet into two parts:
$\tilde{D}_1(s,b)$ which is sent as an {\bf open message}
(after being encrypted by the key)
and $D_2(s,b)$ which is sent as a secure message.
Our scheme uses a separation strategy similar
to \cite{Prabhakaran:Secrecy08} by sending
public and private messages simultaneously.
But in contrast to \cite{Prabhakaran:Secrecy08}, we
exploit the fading channel to secure the key,
and hence, the message.
We now describe the initial key generation and key renewal processes.
For the very first $B$ blocks (the super-block $s=1$),
we generate random key bits, $K(1)$, and then transmit
them from Alice to Bob securely. Utilizing the ergodicity
of the channel, we can transmit a key of an approximate length
$n'B \mathbb{E}[R_k(\hv)]$ bits~\cite{Gopala:On08}.
Then, for any super-block $s>1$, we will use the key
$K(s-1)$ for the one time pad, and also
generate a new key $K(s)$ for the use in the next super-block.
Here, to secure the open packet of block $(s,b)$,
we use $n'R_1(\hv)$ bits from the remaining bits
of the key $K(s-1)$, represented by $\tilde{K}(s,b)$,
to encrypt the data packet
$\tilde{D}_1(s,b)$ using one time pad encryption:
\begin{equation}\label{eq:thm2prfeq3}
D_1(s,b) = \tilde{D}_1(s,b) \oplus \tilde{K}(s,b).
\end{equation}
The encoder will declare an encoding error, if
there are not sufficient key bits left in $K(s-1)$
for the one time pad encryption. To summarize, during the block $(s,b)$, four
messages are {\bf combined together} and sent over the channel:
\begin{enumerate}
\item
$D_1(s,b)$ is mapped into $W_1(s,b) \in \Wc_1 =
\{1,2,\cdots,2^{n'R_1(\hv)}\}$.
\item
$D_2(s,b)$ is mapped into $W_2(s,b) \in  \Wc_2 =
\{1,2,\cdots,2^{n'R_2(\hv)}\}$.
\item
The key bits $D_k(s,b)$ are mapped into $W_k(s,b) \in \Wc_k
=\{1,2,\cdots,2^{n'R_k(\hv)}\}$.
\item
Additional randomization is mapped into $W_x(s,b) \in \Wc_x
=\{1,2,\cdots,2^{n'R_x(\hv)}\}$.

\end{enumerate}

{\bf Codebook Generation and Encoding:}
Our random coding arguments rely on an ensemble 
of codebooks generated according to a zero-mean 
Gaussain distribution with variance $P(\hv)$.
If there are enough number of key bits for the
one time pad scheme (i.e., no encoding error),
the encoder will work as follows.
For a given block $(s,b)$, let $R=\log(1+P(\hv)h_m(s,b)) - \epsilon$.
When $h_m(s,b)\leq h_e(s,b)$, we have one of
$2^{n' (R_1(\hv)+R_x(\hv))}$ open messages,
denoted by the pair $(w_1,w_x)$, to be sent. 
To encode the message $(w_1,w_x)$, the encoder
selects the codeword $X^{n'}(w_1,w_x)$ from 
the chosen codebook. On the other hand,
when $h_m(s,b)\geq h_e(s,b)$, a binning scheme
(see, e.g.,~\cite{Wyner:The75}) is used to send secret
bits over the channel. We first generate a Gaussian codebook
consisting of $2^{n' R}$ codewords, represented by $X^{n'}$,
and then independently assign each of them
to one of $2^{n' (R_k(\hv)+R_2(\hv))}$ bins,
where the bin index is $(w_k,w_2)$,
according to a uniform distribution. This ensures that
any of the sequences are equally likely to be within any
of the bins. Each bin has $2^{n' (R_1(\hv)+R_x(\hv))}$
sequences with codeword index denoted by $(w_1,w_x)$.
Accordingly, a sequence is represented by the tuple of
indices $(w_k,w_2,w_1,w_x)$. To encode a particular
key-message pair, the encoder chooses a codeword indexed by
$(w_1,w_x)$ from the bin indexed by $(w_k,w_2)$, i.e.,
$X^{n'}(w_k,w_2,w_1,w_x)$, and send it
over the channel. We note that $w_x$ is uniformly chosen among $\Wc_x$
and $w_1$ is determined by the data $\tilde{D}_1(s,b)$
and the corresponding key bits of the previous super-block
$\tilde{K}(s,b)$, and hence uniformly distributed over $\Wc_1$.

{\bf Error Analysis:} For each fading block $(s,b)$, we denote the
encoding and decoding error events by
$E_{\textrm{enc}}(s,b)$ and $E_{\textrm{dec}}(s,b)$,
respectively. Then, we write the
error probability at the receiver as follows.
\begin{equation}
P_e^{n'}(s,b) = \Prob\{E_{\textrm{enc}}(s,b)\}
+ \Prob\{E_{\textrm{dec}}(s,b)|E_{\textrm{enc}}^c(s,b)\}
\end{equation}

Since we only impose a constraint on $\mathbb{E}[R_1(\hv)]$
in \eqref{rates-3}, there will be a non-zero
probability that the key bits fall short. In such a case
the encoder will declare an error. Hence, we can write the
following bound.
\begin{equation}\label{eq:thm2prfeq4}
\Prob\{E_{\textrm{enc}}(s,b)\} \leq
\Prob \left\{\sum\limits_{b = 1}^{B}R_1(s,b)
> \sum\limits_{b = 1}^{B}R_k(s-1,b) \right\}
\end{equation}
Here, from the strong law of large numbers and
from \eqref{rates-3}, we see that the right
hand side of \eqref{eq:thm2prfeq4} and hence
$\Prob\{E_{\textrm{enc}}(s,b)\}$ can be arbitrarily made
small as $B\to\infty$.

Now, it remains to show that
$\Prob\{E_{\textrm{dec}}(s,b)|E_{\textrm{enc}}^c(s,b)\}$
can be arbitrarily made small. This follows
as $n'\rightarrow\infty$, by applying the
asymptotic equipartition property and jointly
typical decoding \cite{Cover}. In particular,
for $h_m(s,b)\leq h_e(s,b)$, the messages $w_1$
and $w_x$; and for
$h_m(s,b)\geq h_e(s,b)$, the messages $w_1$, $w_x$,
$w_k$, and $w_2$ can be transmitted reliably.
Furthermore, as $B\rightarrow\infty$, the
average key rate $\mathbb{E}[R_k(\hv)]$
is achievable within any super-block \cite{Gopala:On08}.

{\bf Equivocation Computation:} Here, we show 
that the secrecy condition given by
\eqref{eq:secrecyCondition} is satisfied for each fading
block $(s,b)$, $s>1$. We can write
\begin{eqnarray}
n'R_e(s,b) & \overset{(a)}{=} & H(\tilde{D}_1(s,b), D_2(s,b)|Z^{n},h_m^{n},h_e^{n})\nonumber\\
& = & H(\tilde{D}_1(s,b), D_2(s,b)|Z^{n'}(1,1),Z^{n'}(1,2),\cdots,Z^{n'}(s,b),h_m^{n},h_e^{n})
\nonumber\\
& \overset{(b)}{=} & H(\tilde{D}_1(s,b), D_2(s,b)|Z^{n'}(s,b),Z^{B n'}(s-1),h_m^{n},h_e^{n}) \nonumber\\
& = & H(D_2(s,b)|Z^{n'}(s,b),Z^{B n'}(s-1),h_m^{n},h_e^{n}) \nonumber\\
&& {+}\:
H(\tilde{D}_1(s,b)|D_2(s,b),Z^{n'}(s,b),
Z^{B n'}(s-1),h_m^{n},h_e^{n}) \label{eq:thm2prfeq5}
\end{eqnarray}
where $Z^{B n'}(s-1) = Z^{n'}(s-1,1),Z^{n'}(s-1,2),
\cdots,Z^{n'}(s-1,B)$ is the output of the channel at
the eavesdropper in the previous super-block $s-1$, 
(a) follows from splitting
the data $D(s,b)$ into the two parts
$\tilde{D}_1(s,b)$ and $D_2(s,b)$, and
(b) follows from the independence between block $(s,b)$
and other received signals at the eavesdropper. 
We now focus on the first term in \eqref{eq:thm2prfeq5}.
We note that, in the case where $h_m < h_e$,
no secret bits are sent and hence the first term is zero.
When $h_m > h_e$, in addition to $W_x(s,b)$, we use
the data $D_1(s,b)$ as a randomization signal
to secure the messages $D_2(s,b)$ and $D_k(s,b)$.
In this case, the first term in \eqref{eq:thm2prfeq5}
can be lower bounded by the following two steps. First,
\begin{eqnarray}
\frac{1}{n'}H(D_2(s,b), D_k(s,b)|Z^{n'}(s,b),Z^{B n'}(s-1),h_m^{n},h_e^{n})
&\overset{(a)}{=} & \frac{1}{n'}
H(D_2(s,b), D_k(s,b)|Z^{n'}(s,b),h_m^{n},h_e^{n})\nonumber \\
&\overset{(b)}{\ge} & \frac{1}{n'}
H(D_2(s,b), D_k(s,b)) - \epsilon \nonumber \\
& \overset{(c)}{=} & \frac{1}{n'}
H(D_2(s,b)) + H (D_k(s,b)) - \epsilon
\label{eq:thm2prfeq6}
\end{eqnarray}
where (a) follows from the independence of
$(D_2(s,b), D_k(s,b))$ and the previous super-block,
(b) is a result of using the scheme in \cite{Gopala:On08}
and the results of \cite{Wyner:The75}, i.e., the secrecy
of $D_2(s,b)$ and $D_k(s,b)$, along with the
appropriate choice of the randomization rate such that
$R_1(\hv) + R_x(\hv) = I(X(s,b); Z(s,b))$
and (c) follows from the independence of $D_2(s,b)$
and $D_k(s,b)$. Second, from \eqref{eq:thm2prfeq6}, we have
\begin{eqnarray}
\frac{1}{n'} (H(D_2 | Z^{n'},Z^{B n'}(s-1),h_m^{n},h_e^{n})
+ H(D_k | D_2, Z^{n'},Z^{B n'}(s-1),h_m^{n},h_e^{n}))
& \ge & \frac{1}{n'} H(D_2) + \frac{1}{n'} H (D_k) - \epsilon, \nonumber
\end{eqnarray}
implying
\begin{eqnarray}
\frac{1}{n'} H(D_2 | Z^{n'},Z^{B n'}(s-1),h_m^{n},h_e^{n}) & \ge &
\frac{1}{n'} H(D_2) + \frac{1}{n'}
I(D_k ; D_2, Z^{n'},Z^{B n'}(s-1),h_m^{n},h_e^{n}) - \epsilon \nonumber \\
& \ge & \frac{1}{n'} H(D_2) - \epsilon \label{eq:thm2prfeq7}
\end{eqnarray}
where we have dropped the index $(s,b)$ 
for simplicity of notation and
the last inequality follows from the fact 
that mutual information is non-negative.

The second term in \eqref{eq:thm2prfeq5} is lower bounded as
\begin{eqnarray}
\frac{1}{n'} H(\tilde{D}_1|D_2, Z^{n'},Z^{B n'}(s-1),h_m^{n},h_e^{n})
& = & \frac{1}{n'} H(\tilde{D}_1|Z^{B n'}(s-1),h_m^{n},h_e^{n}) \nonumber \\
&& {-}\: \frac{1}{n'} I(\tilde{D}_1;Z^{n'}, D_2|Z^{B n'}(s-1),h_m^{n},h_e^{n}) \nonumber \\
& = & \frac{1}{n'} H(\tilde{D}_1|h_m^{n},h_e^{n}) \nonumber \\
&& {-}\: \frac{1}{n'} I(\tilde{D}_1;Z^{n'}|Z^{B n'}(s-1),D_2,h_m^{n},h_e^{n}) \label{eq:thm2prfeq8}
\end{eqnarray}
since $\tilde{D}_1$ is independent of $Z^{B n'}(s-1)$ and $D_2$.
The second term in \eqref{eq:thm2prfeq8} is upper bounded as

\begin{eqnarray}
\frac{1}{n'} I(\tilde{D}_1;Z^{n'}|Z^{B n'}(s-1),D_2,h_m^{n},h_e^{n}) & \leq & \frac{1}{n'} I(\tilde{D}_1;Z^{n'},D_1|Z^{B n'}(s-1),D_2,h_m^{n},h_e^{n})\nonumber\\
& = & \frac{1}{n'} H(\tilde{D}_1|Z^{B n'}(s-1),D_2,h_m^{n},h_e^{n}) \nonumber \\
&& {-}\:  \frac{1}{n'} H(\tilde{D}_1|Z^{n'},D_1,Z^{B n'}(s-1),D_2,h_m^{n},h_e^{n})\nonumber\\
& \overset{(a)}{=} & \frac{1}{n'} H(\tilde{D}_1|Z^{B n'}(s-1),h_m^{n},h_e^{n}) \nonumber \\
&& {-}\:  \frac{1}{n'} H(\tilde{D}_1|D_1,Z^{B n'}(s-1),h_m^{n},h_e^{n})\nonumber\\
& = & \frac{1}{n'}
I(\tilde{D}_1;D_1|Z^{B n'}(s-1),h_m^{n},h_e^{n})\nonumber\\
& = & \frac{1}{n'}
H(D_1|Z^{B n'}(s-1),h_m^{n},h_e^{n}) \nonumber \\
&& {-}\:  \frac{1}{n'} H(D_1|\tilde{D}_1,Z^{B n'}(s-1),h_m^{n},h_e^{n})\nonumber\\
& = & \frac{1}{n'} H(\tilde{D}_1 \oplus \tilde{K}|Z^{B n'}(s-1),h_m^{n},h_e^{n}) \nonumber \\
&& {-}\:  \frac{1}{n'} H(\tilde{K}|\tilde{D}_1,Z^{B n'}(s-1),h_m^{n},h_e^{n})\nonumber\\
& \overset{(b)}{\leq} & \frac{1}{n'} H(\tilde{D}_1\oplus \tilde{K})
- \frac{1}{n'} H(\tilde{K}|\tilde{D}_1,Z^{B n'}(s-1),h_m^{n},h_e^{n}) \nonumber \\
& \overset{(c)}{=} & R_1(\hv) - \frac{1}{n'}
H(\tilde{K}|Z^{B n'}(s-1),h_m^{n},h_e^{n}) \label{eq:thm2prfeq9}
\end{eqnarray}
where (a) follows from the conditional 
independence of $\tilde{D}_1$ on  $Z^{n'}$ and $D_2$ given $D_1$
and $Z^{B n'}(s-1)$,
(b) follows from the fact that conditioning
does not increase entropy, and (c) follows from the uniform
distribution of $\tilde{K}$ and the independence of
$\tilde{K}$ and $\tilde{D}_1$ given $Z^{B n'}(s-1)$.

Using the same argument as in \eqref{eq:thm2prfeq6}
and \eqref{eq:thm2prfeq7}, and from \eqref{eq:thm2prfeq4},
it is straightforward to see
\begin{eqnarray}
\frac{1}{n'} H(\tilde{K} | Z^{B n'}(s-1),h_m^{n},h_e^{n})
& \ge & \frac{1}{n'} H(\tilde{K}) - \epsilon \nonumber\\
& = & R_1(\hv) - \epsilon.
\end{eqnarray}
Substituting this in \eqref{eq:thm2prfeq9}
and \eqref{eq:thm2prfeq8}, we get
\begin{eqnarray}
\frac{1}{n'} H(\tilde{D}_1|D_2, Z^{n'},Z^{B n'}(s-1),h_m^{n},h_e^{n}) & \ge &
\frac{1}{n'} H(\tilde{D}_1) - \epsilon \label{eq:thm2prfeq10}
\end{eqnarray}

Finally, combining \eqref{eq:thm2prfeq5}, \eqref{eq:thm2prfeq7},
and \eqref{eq:thm2prfeq10} completes the proof.
\end{IEEEproof}

In the previous result, the achievable rate satisfies the requirements
given by the Definition \ref{def:achievablerate}. Consequently, 
the outage probability is zero with the proposed scheme.
We also remark that, with the above achievability scheme,
the initialization phase is over the first super-block,
during which the data is not transmitted. With a simple
modification, the data can also be transmitted
during the first super-block by sacrificing
the security of \emph{only} the corresponding packets,
which is negligibly small compared to the whole message.

The final step in this section is to establish the asymptotic
optimality of the proposed security scheme in the high
SNR regime. The following result achieves this objective
by showing that the upper and lower bounds of
Theorems~\ref{thm:upperFullCSI} and~\ref{thm:lowerFullCSI}
match in this asymptotic scenario for a wide class of
invertible channels.

\begin{lemma} \label{thm:lemmafull}

In an asymptotic regime of high SNR,
i.e., $\bar{P}\to\infty$, the delay limited secrecy capacity
is given by
\begin{eqnarray} \label{eq:lemmafull}
\lim\limits_{\bar{P} \to \infty} C_{s,d}^{F}
&=& \mathbb{E}_{h_m>h_e}
\left[
\log \left( \frac{h_m}{h_e} \right)
\right],
\end{eqnarray}
assuming that ${\mathbb E}\left[\frac{1}{\min(h_e,h_m)}\right]$
is finite. Moreover, the capacity is achieved by the proposed
one-time pad encryption scheme coupled with the key renewal process.
\end{lemma}

\begin{IEEEproof}
We only need to consider the lower bound as the right
hand side of~\eqref{eq:lemmafull} is the ergodic secrecy capacity in
the high SNR regime, which is by definition an upper bound
on the delay limited secrecy capacity. To this end, in the
proposed scheme, we set $q(\hv) = h_e$ resulting in
$R_2(\hv)=0$. Furthermore, we let
$P(\hv)=\frac{c}{\min(h_e,h_m)}$, where $c$ is a constant,
which is chosen according to the average power constraint.
The achievable rate expression in the high SNR regime is
then given by
\begin{equation}
\lim\limits_{\bar{P} \to \infty}
R_{s,d}^{F} = \lim\limits_{\bar{P} \to \infty}
\min \limits_{\hv}
R_1(\hv),
\end{equation}
where $R_1(\hv)$ is chosen to satisfy
\begin{eqnarray}\label{eq:Lemma3eq2}
\mathbb{E}[R_1(\hv)] &\leq& \mathbb{E}\left[[\log(1+P(\hv)h_m)-
\log(1+P(\hv)h_e)]^+\right]\nonumber\\
R_1(\hv) &\leq& \log(1+c)
\end{eqnarray}

As $\bar{P} \to \infty$, it is easy to see that $c\to \infty$
since ${\mathbb E}\left[\frac{1}{\min(h_e,h_m)}\right]$ is
finite, implying that the second constraint in \eqref{eq:Lemma3eq2}
is loose. Also, it is easy to see that the first constraint
converges to the right hand side of the lemma. Then, by choosing
$R_1(\hv)=\mathbb{E}_{h_m>h_e}
\left[
\log \left( \frac{h_m}{h_e} \right)
\right]$, both constraints of \eqref{eq:Lemma3eq2}
are satisfied and hence the result is proved.
\end{IEEEproof}


\section{Only Main Channel CSI}

In this section we assume that only the legitimate
receiver CSI is available at the transmitter. First, we have
the following upper bound.

\begin{theorem}\label{thm:upperMainCSI}

The delay limited secrecy capacity when only the
legitimate receiver channel state is available at
the transmitter, $C_{s,d}^{M}$,  is upper bounded by
\begin{eqnarray}
C_{s,d}^{M} \leq
\max \limits_{{}^{\quad \: P(h_m)}_{\textrm{s.t. }
{\mathbb E} [P(h_m)]\leq \bar{P} }}
\min
\left\{  R_s^{M} ,  R_d^{M}  \right\}
\end{eqnarray}
\noindent where $R_s^{M}$ and $R_d^{M}$ are given as follows.
\begin{eqnarray}
R_s^{M} &=& \mathbb{E} \left[\log(1+P(h_m)h_m)
- \log(1+P(h_m)h_e)\right]^+\nonumber\\
R_d^{M} &=& \min\limits_{h_m} \log(1 + P(h_m)h_m )\nonumber
\end{eqnarray}
\end{theorem}

\begin{IEEEproof}
The proof follows the same argument as that of
Theorem~\ref{thm:upperFullCSI} with the power
control policy $P(h_m)$.
\end{IEEEproof}

The achievability scheme in this scenario is different from the
previous scenario in two key aspects: 1) the lack of knowledge
about $h_e$ forces us to secure the whole delay sensitive traffic
with the one time pad approach (i.e., setting the rate $R_2(\hv)$ to
zero) and 2) the binning
scheme of the key renewal process must now operate on the level
of the super-block to average-out the fluctuations in $h_e$. On
the other hand, the delay sensitive packet must be decoded after
each block. This makes the use of the delay sensitive 
packet as a randomization signal a rather challenging task. 
Therefore, the achievable rate reported
in the following result is obtained by superimposing the binning
scheme (used to secure the key) on the delay limited traffic
(secured by the key bits sent in the previous super-block).

\begin{theorem}\label{thm:lowerMainCSI}

For any given arbitrarily small $\epsilon$,
the $\epsilon$-delay limited secrecy capacity in the
only main CSI scenario, $C_{s,d}^{M}(\epsilon)$, is lower
bounded by the following $\epsilon$-achievable rate.
\begin{eqnarray}
C_{s,d} ^{M}(\epsilon) \geq R_{s,d}^{M}(\epsilon) =
\max \limits_{{}^{\quad \: P(h_m)}_{\textrm{s.t. }
{\mathbb E} [P(h_m)]\leq \bar{P} }}
\min
\bigg\{
R_s, R_d^{M}
\bigg\},
\end{eqnarray}
where
\begin{eqnarray}
R_s &=& {\mathbb E}
[\log(1+P(h_m)h_m)-R_{s,d}^{M}(\epsilon)
-\log(1+P(h_m)h_e)]^+,\\
R_d^{M} &=&  \min\limits_{h_m} ~ \log(1+P(h_m)h_m).
\end{eqnarray}
\end{theorem}

\begin{IEEEproof}\label{thm:lowerMainCSIproof}
First, fix a power control policy $P(h_m)$.
The achievable rate is then obtained by maximizing over all
power control policies satisfying the average
power constraint. We start by describing our scheme.
We divide the channel uses into super-blocks and
further divide each super-block into blocks such
that the coherence interval is $n'$ symbols as
considered in the proof of Theorem~\ref{thm:lowerFullCSI}.
In this scenario, we utilize the achievable secrecy rate
within a block only for the key generation. That is,
data is transmitted only by using the one-time pad
encryption in contrast to the scheme used in
Theorem~\ref{thm:lowerFullCSI}. Due to the lack of
knowledge of $h_e$, the key is decoded at the end
of each super-block whereas the data packets
are still decoded block by block using the key
sent in the previous super-block. A given message $W\in\{1,2,\cdots,
2^{nR_{s,d}^{M}(\epsilon)}\}$,
is divided into $(S-1)\:B$ data packets,
each represented by $\tilde{D}(s,b)$
for $s\in\{2,\cdots,S\}$ and $b\in\{1,\cdots,B\}$,
where each packet is sent with rate
$R_{s,d}^{M}(\epsilon)$ during
the corresponding block of the channel.
The data packet
$\tilde{D}(s,b)$ is transmitted along with the generated key using
the one-time pad scheme. Initial key generation and key renewal
is similar to the scheme in Theorem~\ref{thm:lowerFullCSI}.
We remark that, similar to Theorem~\ref{thm:lowerFullCSI},
the initialization phase duration
becomes negligible as $S\rightarrow\infty$.

{\bf Codebook Generation and Encoding:} Let $R = \min \{ R_s,R_d^{M}\}$. 
For any given block $(s,b)$, $s>1$, we use the $n'R$
remaining bits from the key $K(s-1)$ and
denote corresponding bits as $\tilde{K}(s,b)$.
These bits are used in a one-time pad scheme to construct
\begin{equation}
D(s,b) = \tilde{D}(s,b) \oplus \tilde{K}(s,b) \label{eq:thm5prfeq1}
\end{equation}
The encrypted bits are then mapped to a message
$w(s,b)\in\{ 1,2,\cdots,2^{n'R} \}$. For the key
renewal process, the binning scheme is
constructed over the super block $s$, as in the
achievable scheme used in~\cite{Gopala:On08},
such that the output bits of the encoder are
divided into $B$ independent blocks each consists of
$n'[\log(1+P(h_m(s,b))h_m(s,b))-R-\epsilon]$ bits
where $b \in \{1,2,\cdots,B\}$ is the coherence interval.
We then combine those bits with the $n' R$ reserved bits for the
encrypted data packet and encode them using a member of the
generated Gaussian codebook ensemble, which has
$2^{n' [\log(1+h_m(s,b) P(h_m(s,b))) -\epsilon]}$ codewords.
The channel input, denoted by $X^{n'}(s,b)$,
corresponding to the message from the code
is sent from the transmitter.

{\bf Error Analysis:} Each codeword is decoded at the end of the block
releasing the delay sensitive packet. Following the
same argument used in proof of Theorem \ref{thm:lowerFullCSI},
$P_e^{n'}(s,b)$ can be made arbitrarily small as
$n' \rightarrow \infty$ for each $(s,b)$. The key bits are
decoded at the end of the binning codeword (i.e., super block)
following the same argument used in \cite{Gopala:On08}.
Therefore, as $n' \rightarrow \infty$ and $B \rightarrow \infty$,
the proposed key rate is achievable, where the encrypted
data bits are not used as a part of the randomization message.

{\bf Equivocation Computation:} We will show that, 
for the given $\epsilon$ (can be arbitrarily
small) and for any given $\epsilon'>0$,
$$P_{out}(s,b,R_{s,d}(\epsilon),\epsilon') \leq \epsilon$$
with the proposed scheme. The equivocation per block for
arbitrary chosen block $(s,b)$ is given by

\begin{eqnarray}
n'R_e(s,b) & = & H(\tilde{D}(s,b)|Z^{n},h_m^{n},h_e^{n})\nonumber\\
& = & H(\tilde{D}|Z^{n'}(1,1),Z^{n'}(1,2),\cdots,Z^{n'}(S,B),h_m^{n},h_e^{n})\nonumber\\
& \overset{(a)}{=} & H(\tilde{D}|Z^{n'}(s,b),Z^{B n'}(s-1),h_m^{n},h_e^{n})\nonumber \\
& = & H(\tilde{D}|Z^{B n'}(s-1),h_m^{n},h_e^{n})
- I(\tilde{D};Z^{n'}|Z^{B n'}(s-1),h_m^{n},h_e^{n}) \nonumber \\
& \overset{(b)}{=} & H(\tilde{D}|h_m^{n},h_e^{n})
- I(\tilde{D};Z^{n'}|Z^{B n'}(s-1),h_m^{n},h_e^{n}) \nonumber \\
& \geq & H(\tilde{D}|h_m^{n},h_e^{n})
- I(\tilde{D};Z^{n'},D|Z^{B n'}(s-1),h_m^{n},h_e^{n})\nonumber\\
& = & H(\tilde{D}|h_m^{n},h_e^{n})
- H(\tilde{D}|Z^{B n'}(s-1),h_m^{n},h_e^{n})
+ H(\tilde{D}|Z^{n'},D,Z^{B n'}(s-1),h_m^{n},h_e^{n})\nonumber\\
& \overset{(c)}{=} & H(\tilde{D}|h_m^{n},h_e^{n})
- H(\tilde{D}|Z^{B n'}(s-1),h_m^{n},h_e^{n})
+ H(\tilde{D}|D,Z^{Bn'}(s-1),h_m^{n},h_e^{n})\nonumber\\
& = & H(\tilde{D}|h_m^{n},h_e^{n})
- I(\tilde{D};D|Z^{Bn'}(s-1),h_m^{n},h_e^{n})\nonumber\\
& = & H(\tilde{D}|h_m^{n},h_e^{n})
- H(D|Z^{B n'}(s-1),h_m^{n},h_e^{n})
+ H(D|\tilde{D},Z^{Bn'}(s-1),h_m^{n},h_e^{n})\nonumber\\
& \geq & H(\tilde{D}|h_m^{n},h_e^{n})
- H(\tilde{D} \oplus \tilde{K})
+ H(\tilde{K}|\tilde{D},Z^{Bn'}(s-1),h_m^{n},h_e^{n})\nonumber\\
& \overset{(d)}{=} & H(\tilde{D}|h_m^{n},h_e^{n})
- H(\tilde{D} \oplus \tilde{K})
+ H(\tilde{K}|Z^{Bn'}(s-1),h_m^{n},h_e^{n})
\label{eq:thm5prfeq3}
\end{eqnarray}
where $Z^{B n'}(s-1)
= Z^{n'}(s-1,1),Z^{n'}(s-1,2),\cdots,Z^{n'}(s-1,B)$
is the output of the channel in the previous super-block $s-1$, 
the index $(s,b)$ is omitted in $\tilde{D}(s,b)$ and $\tilde{K}(s,b)$,
(a) follows from the independence between block
$(s,b)$ and other transmissions, (b) follows from the independence of $\tilde{D}$
and $Z^{B n'}(s-1)$, (c) follows from the independence
between $\tilde{D}$ and $Z^{n'}$ given $D$ and
$Z^{B n'}(s-1)$,
and (d) follows from the independence
of $\tilde{K}$ and $\tilde{D}$  given $Z^{Bn'}(s-1)$.

It remains to bound the last term in
\eqref{eq:thm5prfeq3}. Here, if we satisfy
\begin{equation}
 \frac{1}{n'} H(\tilde{K}|Z^{Bn'},h_m^n,h_e^n)
 \geq \frac{1}{n'}  H(\tilde{K}) - \epsilon', \label{eq:thm5prfeq4}
\end{equation}
we have, from \eqref{eq:thm5prfeq3}, that
$$R_e(s,b) \geq R_{s,d} - \epsilon'.$$
Therefore, the secrecy outage event happens once
\eqref{eq:thm5prfeq4} is not satisfied with the
given $\epsilon'$. We denote this event as follows.
\begin{equation}\label{eq:thm5prfeq5}
\mathcal{O}(\epsilon')
\triangleq \left\{
\frac{1}{n'} H(\tilde{K}|Z^{Bn'},h_m^n,h_e^n)
< \frac{1}{n'}  H(\tilde{K}) - \epsilon'\right\}
\end{equation}
Consequently, we will use the bound
\begin{equation}\label{eq:thm5prfeq6}
P_{out}(s,b,R_{s,d}(\epsilon),\epsilon')
\leq
\Prob \left\{\mathcal{O}(\epsilon')\right\}
\end{equation}
in order to show that the outage probability
can be made less than $\epsilon$.

Following the argument given in~\cite{Gopala:On08},
one can see that the following key rate can be
achieved with perfect secrecy
(as $n'\rightarrow\infty$ and $B\rightarrow\infty$).
$$ R_s= {\mathbb E}[
\log(1+P(h_m)h_m)-R
-\log(1+P(h_m)h_e)
]^+ $$
with
\begin{equation}
\frac{1}{n'B} H(K(s-1)|Z^{B n'}(s-1),h_m^{n},h_e^{n}) \geq
\frac{1}{n'B} H(K(s-1)) - \epsilon_1 \label{eq:thm5prfeq2}
\end{equation}
where $Z^{B n'}(s-1)$ is the received signal
by Eve for the super-block $s-1$ and $\epsilon_1 > 0$
is arbitrarily small as $n',B \rightarrow \infty$.
Here, we denote the number of blocks within the super block $s-1$
for which the event $\mathcal{O}(\epsilon')$ holds
as $\beta$. Then,
from \eqref{eq:thm5prfeq5} and \eqref{eq:thm5prfeq2},
we conclude that $n' \epsilon' \beta \leq n'B \epsilon_1$,
which further implies $$\Prob\{\mathcal{O}(\epsilon')\} =
\lim_{B\to\infty} \frac{\beta}{B}
\leq \lim_{B\to\infty} \frac{\epsilon_1}{\epsilon'}.$$
At this point, as $\epsilon_1$ can be arbitrarily made small
as $n',B\to\infty$, we conclude from \eqref{eq:thm5prfeq5}
that, for any given arbitrarily
small $\epsilon'$ and $\epsilon$
$$P_{out}(s,b,R_{s,d}(\epsilon),\epsilon') \leq \epsilon$$
for sufficiently large $n'$ and $B$.
\end{IEEEproof}

We note that, when only the main CSI is available, we followed the
$\epsilon$-achievability notion given by definition
\ref{def:epsachievablerate} with some arbitrarily small $\epsilon$,
which means that the above claimed rate is achievable with perfect
secrecy for every realization of the channel except for a
subset whose probability can be arbitrarily made small.

\begin{figure}[thb]
   \centering
   \includegraphics[width=0.8\columnwidth]{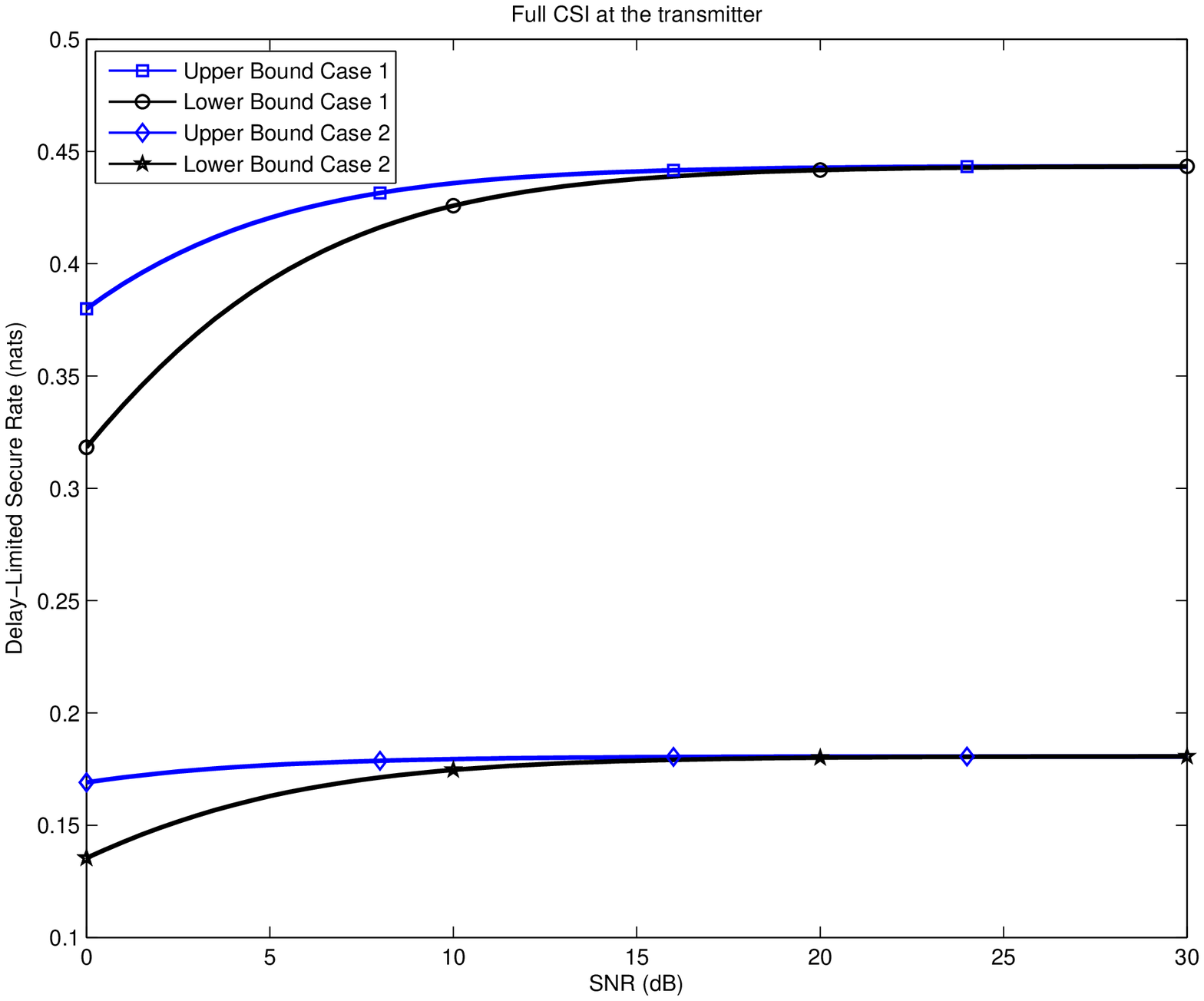}
    \caption{Simulation results for the Full CSI.
    ${\mathbb E}[h_e] = {\mathbb E}[h_m]$ in case 1, and
    ${\mathbb E}[h_e] = 2~{\mathbb E}[h_m]$ in case 2.}
    \label{fig:FullCSI}
\end{figure}

\begin{figure}[thb]
   \centering
   \includegraphics[width=0.8\columnwidth]{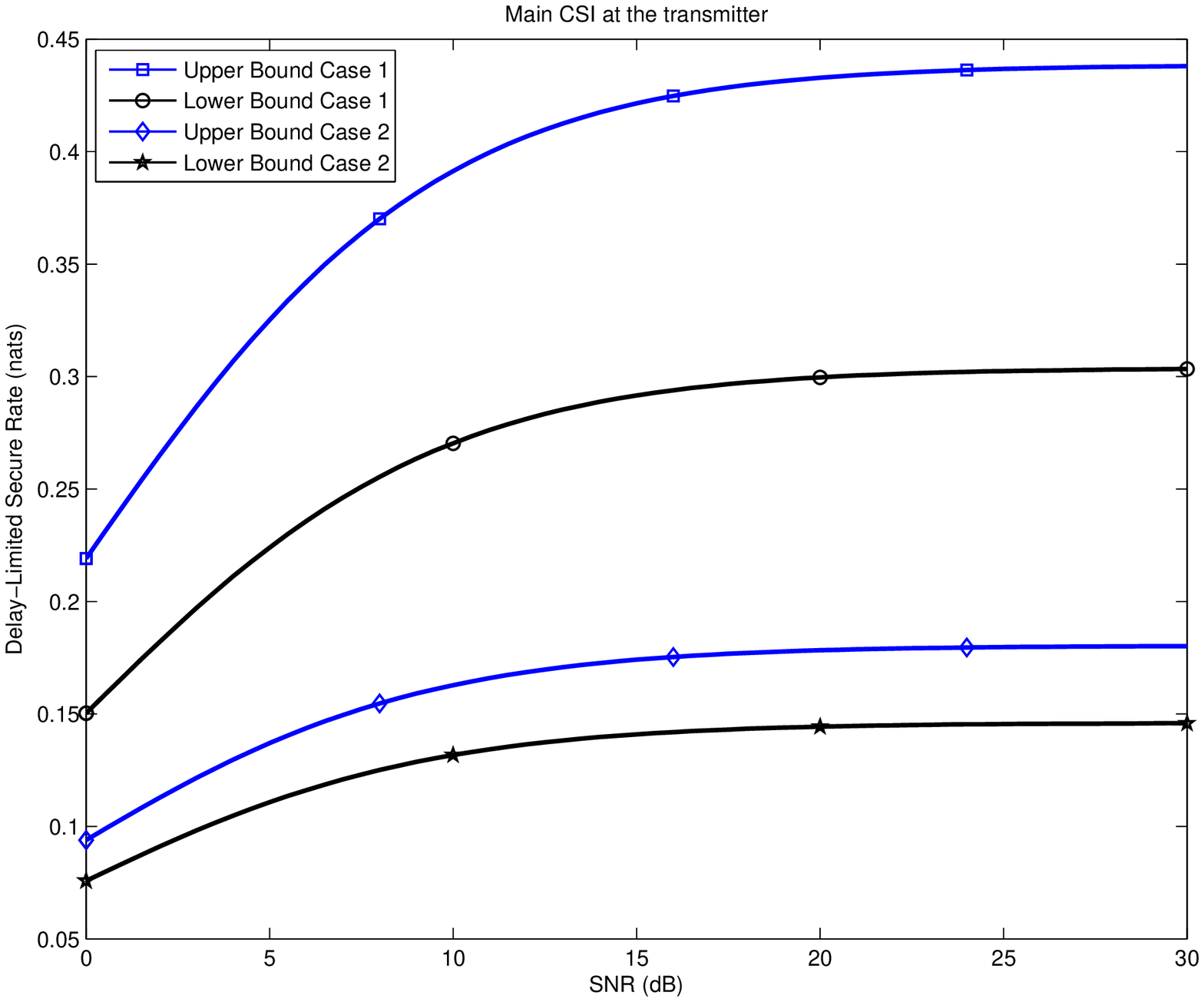}
    \caption{Simulation results for the Main CSI.
    ${\mathbb E}[h_e] = {\mathbb E}[h_m]$ in case 1, and
    ${\mathbb E}[h_e] = 2~{\mathbb E}[h_m]$ in case 2.}
    \label{fig:MainCSI}
\end{figure}

Finally, we report numerical results that 
validate our theoretical claims.
In the full CSI case, we set $q(\hv)=h_e$
(hence $R_2(\hv)=0$) and use channel inversion power control
policy for the achievable rate. We first set both $h_m$ and $h_e$
to be independent and identically distributed
Chi-Square random variables with four degrees of freedom.
Remarkably, as shown in the upper two curves in Fig.~\ref{fig:FullCSI},
even with these, in general suboptimal, choices of
$q(\hv)$ and $P(\hv)$, the lower and upper bounds
coincide in the high SNR regime. The same trend is 
observed in the lower two curves of the figure corresponding to
the case with ${\mathbb E}[h_e]
= 2{\mathbb E}[h_m]$. Figure~\ref{fig:MainCSI} 
corresponds to the case where only
the main channel CSI is available at the transmitter. 
Here, the channel inversion power control policy is used
for both the upper and lower bounds.
The achievability of a non-zero delay limited rate 
is evident even for the case in which the eavesdropper
channel is better than the main channel on the average.


\section{Conclusions}

We have studied the delay limited secrecy capacity of the slow-fading
channel under different assumptions on the transmitter CSI.
Our achievability arguments are based on a novel two-stage
scheme that allows for overcoming the secrecy outage phenomenon
for a wide class of channels. The scheme is based on sharing
{\em a delay tolerant} private key, using random binning, and then
using the key to encrypt the {\em the delay sensitive} packets in a one
time pad format. For the full CSI case, our scheme is shown to
be asymptotically optimal, i.e., at high SNR regime, for many relevant
channel distributions. When only the main channel CSI is available,
the two-stage scheme achieves a non-zero
delay-limited secure rate, with high probability, for invertible channels. Finally, one can easily
identify several avenues for future works. For example,
1) obtaining sharp capacity results for finite values of SNR,
2) extending the results to multiuser scenarios,
3) characterizing the optimal power control policies, and
4) extending the framework to bursty traffic by allowing for buffer delays.



\end{document}